\documentclass[12pt,aps,prb,amsmath,amssymb]{revtex4-2}%notitlepage,,preprint

\usepackage{bm}
\usepackage{graphicx}
\usepackage{xcolor}
\usepackage{transparent}

\usepackage{times}
\usepackage{mathptmx}

\usepackage{subcaption}

\newcommand{\wh}{\widehat}
\newcommand{\wt}{\widetilde}

\newcommand{\ave}[1]{{\left<#1\right>}}

\newcommand{\abs}[1]{{\left|#1\right|}}

\newcommand{\rmd}{\text{d}}

\newcommand{\taud}{\ensuremath{\tau_\text{d}}}

\newcommand{\tauw}{\ensuremath{\tau_\text{w}}}
\newcommand{\taumax}{\ensuremath{\tau_\text{max}}}

\newcommand{\tlag}{\triangle_t}
\newcommand{\xlag}{\triangle_x}
\newcommand{\ylag}{\triangle_y}

\newcommand{\ellx}{\ell_\shortparallel}
\newcommand{\elly}{\ell_\perp}
\newcommand{\tx}{\theta_\shortparallel}
\newcommand{\ty}{\theta_\perp}
\newcommand{\tilt}{\alpha}
\newcommand{\vdir}{\beta}
\newcommand{\dt}{\Delta_t}
\newcommand{\dx}{\Delta_x}
\newcommand{\dy}{\Delta_y}

\newcommand{\yk}{\upsilon_k}
\newcommand{\tk}{s_k}

\newcommand{\aave}{\ensuremath{\ave{a}}}

\newcommand{\Phirms}{\ensuremath{\Phi}_\text{rms}}

\newcommand{\MSE}{\text{MSE}}

\newcommand{\Eqref}[1]{Eq.~\eqref{#1}}
\newcommand{\Eqsref}[1]{Eqs.~\eqref{#1}}
\newcommand{\Figref}[1]{Fig.~\ref{#1}}

\newcommand{\Secref}[1]{Sec.~\ref{#1}}
\newcommand{\Secsref}[1]{Secs.~\ref{#1}}
\newcommand{\Appref}[1]{Appendix~\ref{#1}}

\begin{document}

\title{Time delay velocity estimation from a superposition of localized and uncorrelated pulses}

\author{J.~M.~Losada}
\email{juan.m.losada@uit.no}

\author{O.~E.~Garcia}
\email{odd.erik.garcia@uit.no}

\affiliation{Department of Physics and Technology, UiT The Arctic University of Norway, N-9037 Troms{\o}, Norway}

\date{\today}

\begin{abstract}
This study investigates a novel method for estimating two-dimensional velocities using coarse-grained imaging data, which is particularly relevant for applications in plasma diagnostics. The method utilizes measurements from three non-collinear points and is derived from a stochastic model that describes the propagation of uncorrelated pulses through two-dimensional space. We demonstrate that the method provides exact time delay estimates when applied to a superposition of Gaussian structures and remains accurate for various other pulse functions. Through extensive numerical simulations, we evaluate the method's performance under variations in signal duration, pulse overlap, spatial and temporal resolution, and the presence of additive noise. Additionally, we investigate the impact of temporal pulse evolution due to linear damping and explore the so-called barberpole effect, which occurs with elongated and tilted structures. Although the method is susceptible to the barberpole effect, we analytically demonstrate that this effect does not occur when the elongated structures propagate parallel to one of their axes, and we establish bounds for the associated errors. We propose a series of safeguards to anticipate the applicability of the velocity estimation method, considering factors such as signal length, number of pulses, temporal and spatial resolution, signal-to-noise ratio, and pulse size. However, these safeguards do not ensure applicability in cases involving the barberpole effect or correlations between amplitudes and velocities. Overall, our findings provide a comprehensive and robust framework for accurate two-dimensional velocity estimation, enhancing the capabilities of fusion plasma diagnostics and potentially benefiting other fields requiring precise motion analysis.
\end{abstract}

\maketitle

\section{Introduction}

The scrape-off layer (SOL), the outermost region of plasma in a magnetic confinement device, plays a critical role in determining the overall performance and stability of fusion reactors \cite{labombard_cross-field_2000,vianello_scrape-off_2020,stagni_dependence_2022}. In this region, where the plasma interacts with the surrounding materials, various turbulence phenomena lead to the formation of coherent structures known as filaments or blobs. These blobs, characterized by their high-pressure nature and radial motion, dominate the cross-field particle and heat transport \cite{dippolito_blob_2004,zweben_edge_2007}. Understanding the dynamics of plasma blobs is key for developing plasma regimes compatible with continuous, steady-state operation of fusion reactors.

Stochastic modelling has become a commonly used tool to describe the complex dynamics in the SOL \cite{garcia_stochastic_2012,kube_convergence_2015,garcia_stochastic_2016,garcia_auto-correlation_2017,theodorsen_statistical_2017,militello_scrape_2016,militello_relation_2016,theodorsen_level_2018,theodorsen_probability_2018,ahmed_reconstruction_2023}. These models have demonstrated significant utility in accurately capturing and interpreting experimental data, providing a framework to describe the statistical properties and turbulent nature of the plasma dynamics observed in the SOL \cite{garcia_burst_2013,garcia_intermittent_2013,garcia_intermittent_2015,kube_fluctuation_2016,garcia_sol_2017,theodorsen_relationship_2017,walkden_interpretation_2017,theodorsen_universality_2018,kube_intermittent_2018,bencze_characterization_2019,kuang_plasma_2019,kube_comparison_2020,zweben_temporal_2022, zurita_stochastic_2022,ahmed_strongly_2023}. In this work, we extend a well-studied stochastic model to two spatial dimensions, capturing the more complex interactions and dynamics present in real experimental conditions. This is equivalent to the model presented in Ref.~\cite{militello_two-dimensional_2018}. This model facilitates better descriptions of plasma behaviour and supports the development of more effective analysis tools for plasma diagnostics.

In particular, we are interested in developing techniques for velocity estimation on coarse-grained imaging data, such as avalanche photodiode gas-puff imaging (GPI) \cite{cziegler_experimental_2010,zweben_invited_2017,offeddu_gas_2022, terry_realization_2024} or beam emission spectroscopy \cite{mckee_beam_1999, wang_beam_2023}. This diagnostic is characterized by its high temporal resolution but relatively low spatial resolution. Some traditional velocimetry techniques are based on two-point time delay estimation \cite{grulke_radially_2006, ghim_measurement_2012, cziegler_fluctuating_2013, enters_testing_2023, diab_role_2024}. These techniques lead to systematic errors when the velocity of the flow is not aligned with the two points used for the estimation \cite{brotankova_measurement_2009, fedorczak_physical_2012,brandt_investigating_2016,sierchio_comparison_2016}. More advanced methods based on two-dimensional cross-correlation functions are typically applied when higher spatial resolution is available \cite{terry_velocity_2005, zweben_edge_2015, zweben_estimate_2011, lampert_novel_2021, agostini_edge_2011}. Other methods have been studied based on dynamic programming \cite{kriete_extracting_2018, enters_testing_2023} and blob tracking \cite{kube_blob_2013, zweben_blob_2016, offeddu_cross_2022}. These methods are not applicable to the low spatial resolutions of avalanche photodiode GPI \cite{sierchio_comparison_2016}. A three-point velocity estimation method has been demonstrated in the case of propagating waves \cite{brandt_investigating_2016, losada_three-point_2024}.

In this work, we apply a stochastic model to analytically demonstrate that the three-point velocity estimation technique gives exact results for a superposition of uncorrelated, Gaussian structures. We test this technique on synthetic data to study its limits of applicability and the effect of signal duration, degree of pulse overlap, spatial and temporal resolution, broad distributions of velocities, noise, elongated and tilted pulses, time-dependent amplitudes and correlations between pulse velocities and amplitudes. We show that this technique is not subject to the dimensionality limitation of two-point techniques and that it is applicable to coarse spatial resolutions, making it a complementary tool to analysis methods designed for high spatial resolution diagnostics.

Another issue many velocity estimation techniques are subject to is the so-called barberpole effect. This effect appears when a structure such as a blob or a wavefront propagates in a direction not aligned with its normal. This effect can, for example, lead to spurious horizontal velocities even if the structures propagate only vertically \cite{fedorczak_physical_2012}. The proposed three-point method is subject to the barberpole effect, however, we show that the errors associated with this are bounded and lower than those under traditional two-point techniques.

This paper is organized as follows. In \Secref{sec.model} the two-dimensional stochastic model describing the fluctuations as a superposition of propagating and uncorrelated pulses including tilting is introduced. In \Secref{sec.estimation} we derive the three-point velocity estimation method within the statistical framework provided by the stochastic model. In \Secref{sec.test} we test the method on synthetic data for a wide variety of scenarios. A discussion of the method and conclusions of these investigations are presented in \Secref{sec.dc}. Finally, Python implementations of both the velocity estimation methods and the synthetic data generation described in this paper are openly available on GitHub \cite{losada_velocity_estimation, cosmo_blobmodel}.

\section{A superposition of pulses in two dimensions}\label{sec.model}
\newcommand{\rate}{\mathcal{R}}

In this section, we describe a stochastic model describing the propagation of uncorrelated structures in two dimensions. This model is similar to that described in Ref.~\cite{militello_two-dimensional_2018}. We consider a process given by a superposition of pulses
\begin{equation}\label{eq.2d}
    \Phi_K(x, y, t) = \sum_{k=1}^{K} \phi_k(x, y-\yk, t-\tk).
\end{equation}
Here $x$ and $y$ are spatial coordinates, which we will refer to as the horizontal and vertical coordinates, respectively. Each pulse $\phi_k$ arrives at $x=0$, $y=\yk$ at time $t=\tk$, where $\yk$ and $\tk$ are random variables, uniformly distributed on $[-L/2,L/2]$ and $[-T/2, T/2]$, respectively. $K$ is the total number of pulses in a given realization of the process.

In order to keep the average waiting time between pulse arrivals constant, we assume that the number of pulses scales with the process duration $T$ and the vertical domain size $L$ as $K = TL/\rate$, where $\rate$ is a rate parameter in units of distance times time. We can compute the expected value of any function $f_k(\tk, \yk, \ldots)$ where the dots indicate other possible random variables such as sizes or shape parameters. We neglect end effects by taking the limits $T, L \rightarrow \infty$, which give the average
\begin{equation}\label{eq.avg.recepie}
    \ave{ \sum_{k=1}^K f_k(\yk, \tk, \ldots)} = \frac{TL}{\rate} \int_{-\infty}^{\infty} \frac{\rmd \tk}{T} \int_{-\infty}^{\infty} \frac{\rmd \yk}{L} \ave{f_k(\yk, \tk, \ldots)} = \frac{\ave{f_k(\yk, \tk, \ldots)}}{\rate}.
\end{equation}
The average $\ave{\cdot}$ is to be performed over $\tk$, $\yk$ and any other random variables.

The parameter $\rate$ is the inverse of the density of pulses appearing in the $(y-t)$-space. In particular, $\tauw=\rate/L$ is the average waiting time between pulses arriving at any vertical coordinate. For later reference, we introduce the normalized process
\begin{equation}\label{eq.phi.norm}
    \wt{\Phi}(x,y,t) = \frac{\Phi(x,y,t)-\ave{\Phi}}{\Phirms},
\end{equation}
where $\ave{\Phi}$ and $\Phirms$ denote the mean value of the process and the root mean square value, which both will be later shown to be independent of the space and time coordinates.

\subsection{Pulse evolution}

\begin{figure*}[tb]
    \centering
    \begin{subfigure}[b]{0.45\textwidth}
    \def\svgwidth{.9\columnwidth}
    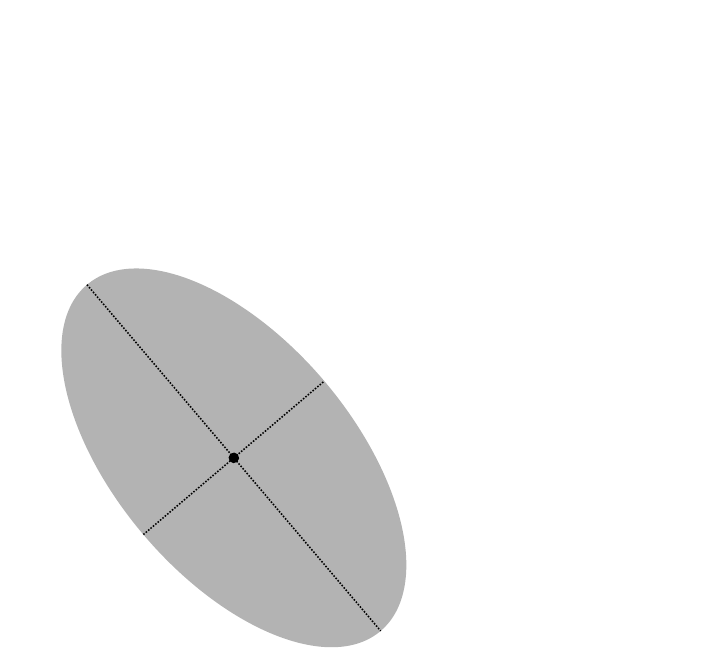
    \end{subfigure}
    \hfill
    \begin{subfigure}[b]{0.45\textwidth}
        \centering
        \def\svgwidth{.9\columnwidth}
        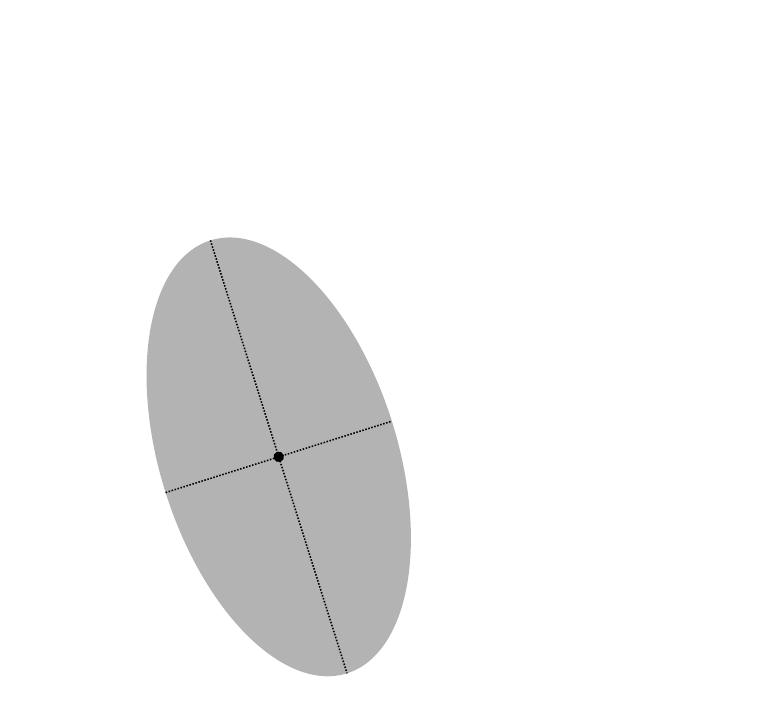
    \end{subfigure}
    \caption{Left: Tilted pulse, where the pulse orientation $\tilt$ does not correspond to the velocity direction, $\tilt\neq\vdir=\text{arctan}(w/v)$. Right: Aligned pulse, where the pulse orientation $\tilt$ is aligned with the velocity direction, $\tilt = \vdir=\text{arctan}(w/v)$.}
    \label{fig.parameters}
\end{figure*}

We assume that each pulse $\phi_k$ has an amplitude $a_k$ and moves with constant velocity $u=\sqrt{v^2+w^2}$, where $v$ and $w$ are the horizontal and vertical velocity components, respectively. Each pulse has a tilt angle $\tilt$ with respect to the horizontal axis. We denote $\tx$ and $\ty$ as the normalized coordinates along a frame aligned and traveling with the pulse. An illustration is provided in \Figref{fig.parameters}. We write the pulse as
\begin{equation}\label{eq.fil_2d}
    \phi_k(x,y,t) = a_k \varphi\left( \tx(x,y,t), \ty(x,y,t)\right),
\end{equation}
where
\begin{subequations}\label{eq.t.both}
\begin{gather}
    \tx(x,y,t) = \frac{(x-vt) \cos \tilt + (y-wt) \sin \tilt}{\ellx}, \label{eq.tx}
    \\
    \ty(x,y,t) = \frac{-(x-vt)\sin\tilt + (y-wt)\cos \tilt}{\elly}. \label{eq.ty}
\end{gather}
\end{subequations}
This describes a pulse centered at $(x,y)=(0,0)$ at time $t=0$ propagating with velocity components $v$ and $w$ in the $x$- and $y$-direction, respectively. The pulse sizes along its axes are denoted by $\ellx$ and $\elly$. Equations (\ref{eq.t.both}) can be written in terms of the velocity angle $\vdir=\arctan (w/v)$ and the velocity magnitude $u$ as
\begin{subequations}
\begin{gather}
    \tx(x,y,t) = \frac{x \cos \tilt + y \sin \tilt}{\ellx} - \frac{ut}{\ellx} \cos (\tilt - \vdir), \label{eq.tx.b}
    \\
    \ty(x,y,t) = \frac{-x \sin\tilt + y \cos \tilt}{\elly} + \frac{ut}{\elly} \sin (\tilt - \vdir), \label{eq.ty.b}
\end{gather}
\end{subequations}
showing that the perpendicular coordinate $\ty(x,y,t)$ becomes time-independent when $\tilt=\vdir$.  Moreover, we assume the pulse function $\varphi$ to be separable,
\begin{equation}\label{eq.ps.general}
    \varphi(\tx, \ty) = \varphi_\shortparallel(\tx) \varphi_\perp(\ty).
\end{equation}
In the case $\tilt=\vdir$, see \Figref{fig.parameters}, the pulse is tilted in the direction it propagates. In this case, $\ellx$ is the pulse size observed by a fixed measurement point, which would only record the pulse variation from $\tx$. The case $\tilt=\vdir$ will be taken as the base case from which the estimation method is derived. Deviations from this will lead to the barberpole effect and will be considered in \Secref{sec.tilt}.

The pulse autocorrelation function is defined as
\begin{subequations}\label{eq.ps.acf}
\begin{equation}
    \rho_\shortparallel(\triangle_\theta) = \int_{-\infty}^{\infty} \rmd \theta \varphi_\shortparallel(\theta) \varphi_\shortparallel(\theta + \triangle_\theta),
\end{equation}
\begin{equation}
    \rho_\perp(\triangle_\theta) = \int_{-\infty}^{\infty} \rmd \theta \varphi_\perp(\theta) \varphi_\perp(\theta + \triangle_\theta),
\end{equation}
\end{subequations}
and the moments of the pulse function, $I_{\shortparallel,n}$ and $I_{\perp,n}$, are defined as
\begin{subequations}\label{eq.ps.mom}
\begin{equation}
    I_{\shortparallel,n} = \int_{-\infty}^{\infty} \rmd \theta \varphi_\shortparallel^n(\theta),
\end{equation}
\begin{equation}
    I_{\perp,n} = \int_{-\infty}^{\infty} \rmd \theta \varphi_\perp^n(\theta).
\end{equation}
\end{subequations}
In the following, we will assume the pulse function to be Gaussian,
\begin{equation}\label{eq.gauss}
    \varphi_\shortparallel(\theta) = \varphi_\perp(\theta) = \frac{1}{\sqrt{\pi}} \exp (-\theta^2),
\end{equation}
in which case the pulse auto-correlation function becomes
\begin{equation}\label{eq.gauss.acf}
    \rho_\shortparallel(\triangle_\theta) = \rho_\perp(\triangle_\theta) = \frac{1}{\sqrt{2\pi}}\exp \left( - \frac{\triangle_\theta^2}{2} \right).
\end{equation}
Other pulse functions are considered in \Appref{app.ps}.

In the following, we assume the pulse sizes $\ellx$ and $\elly$ and velocities $v$ and $w$ to be the same for all pulses unless otherwise stated. This is straightforward to generalize to a distribution of the pulse parameters, as any statistical quantity obtained for the degenerate case can be generalized to a distribution by averaging over the relevant random variables.

\subsection{Mean and variance}

The average of the process can now be computed using \Eqref{eq.avg.recepie},
\begin{equation}
    \ave{\Phi} = \frac{1}{\rate} \ave{\phi_k(x, y-\yk, t-\tk)}.
\end{equation}
Writing it in terms of the pulse function in \Eqref{eq.fil_2d} and using that the amplitudes $a_k$ are independent of arrival times $\tk$ and arrival positions $\yk$ we have
\begin{align}
    \ave{\Phi} = \frac{\ave{a_k}}{\rate}\int_{-\infty}^{\infty} \text{d}\tk \int_{-\infty}^{\infty} \text{d}\yk\, \varphi\left( \tx(x,y-\yk,t-\tk), \ty(x,y-\yk,t-\tk)\right)
\end{align}
We proceed by changing the integration variables to $\tx$ and $\ty$. The Jacobian determinant of the variable transformation is given by $v/\ellx\elly$ and thus the mean value of the process is given by
\begin{equation}\label{eq.ave}
    \ave{\Phi} = \frac{\ave{a} \ellx \elly }{\rate v} I_{\shortparallel,1} I_{\perp,1}.
\end{equation}
Note that \Eqref{eq.ave} is independent of the vertical velocity component $w$. Following a similar procedure we can compute the variance of the process to obtain
\begin{equation}\label{eq.variance}
    \Phirms^2 = \ave{(\Phi - \ave{\Phi})^2} = \frac{\langle {a^2} \rangle \ellx\elly}{\rate v}I_{\shortparallel,2} I_{\perp,2}.
\end{equation}
This has the same velocity dependence as the mean value. For higher velocities, both quantities become smaller, as the pulses spend less time at any given location. Larger pulses will lead to larger mean and variance values, as expected. Increasing the rate parameter $\rate$, which is equivalent to reducing the number of pulses, leads to a smaller mean value and variance. In contrast, the relative fluctuation level is given by
\begin{equation}
    \frac{\Phirms}{\ave{\Phi}} = \sqrt{\frac{\rate v}{\ellx \elly}} \sqrt{\frac{\langle {a^2} \rangle }{\ave{a}^2}} \frac{\sqrt{I_{\shortparallel, 2} I_{\perp, 2}}}{I_{\shortparallel, 1} I_{\perp, 1}}.
\end{equation}
This shows that, for fixed pulse sizes and shapes, the relative fluctuation level increases with the rate parameter $\rate$ and the horizontal velocity component $v$.

\subsection{Cross-correlation function}\label{sec.acf}

The cross-correlation function of the process for a horizontal lag $\xlag$, vertical lag $\ylag$ and temporal lag $\tlag$, is defined as
\begin{equation}
    R_\Phi(\xlag, \ylag, \tlag) = \ave{\Phi(x,y,t) \Phi(x+\xlag,y+\ylag,t + \tlag)}.
\end{equation}
In \Appref{app.ccf}, we show that in the case of Gaussian pulses, the cross-correlation function of the normalized process defined by \Eqref{eq.phi.norm} becomes
\begin{equation}\label{eq.ccf.2g}
    R_{\wt{\Phi}}(\xlag, \ylag, \tlag) = \exp \left( -\frac{\triangle_{\tx}^2 + \triangle_{\ty}^2}{2} \right),
\end{equation}
where
\begin{subequations}\label{eq.txy}
\begin{equation}
    \triangle_{\tx} = \frac{\xlag - v \tlag}{\ellx} \cos \tilt + \frac{\ylag - w \tlag}{\ellx} \sin \tilt, \\
\end{equation}
\begin{equation}
    \triangle_{\ty} = -\frac{\xlag - v \tlag}{\elly} \sin \tilt + \frac{\ylag - w \tlag}{\elly} \cos \tilt.
\end{equation}
\end{subequations}
This provides an analytical expression for the cross-correlation function in the case of Gaussian pulses. In \Secref{sec.estimation}, we will employ this expression to justify the three-point velocity estimation method. Equations~(\ref{eq.txy}) can be rewritten in terms of $v=u\cos\vdir$ and $w=u\sin\vdir$ as
\begin{subequations}\label{eq.txy.ubeta}
\begin{equation}
    \triangle_{\tx} = \frac{\xlag \cos\tilt + \ylag\sin\tilt}{\ellx} - \frac{u\tlag}{\ellx} \cos (\tilt-\vdir), \\
\end{equation}
\begin{equation}
    \triangle_{\ty} = \frac{-\xlag \sin\tilt + \ylag\cos\tilt}{\elly} + \frac{u\tlag}{\elly} \sin (\tilt-\vdir), \\
\end{equation}
\end{subequations}
Therefore, for the base case in which the pulse is tilted along the direction of propagation, $\tilt=\vdir$, we have that the dependence on the time lag $\tlag$ appears only on the parallel component $\triangle_{\tx}$.

% A realization of the model has been drawn to test the validity of \Eqref{eq.ccf.2g}. The details of the realization are discussed in \Secref{sec.estimation}. The signal resulting from the realization is measured at points $P_0=(0, 0)$, $P_x=(1, 0)$ and $P_y=(0, 1)$ where distances are measured in units of sizes $\ellx = \elly$. $\wh{R}_x$ and $\wh{R}_y$ are the cross-correlation functions estimated between the signal measured at points $P_0$ and $P_x$, and $P_0$ and $P_y$ respectively. The results are shown in \Figref{fig.ccf}. Dashed lines are the analytical expression \Eqref{eq.ccf.2g}. The velocities employed for the realization are $v=2w$.
% [Should this figure and discussion be included? Seems a bit too trivial.]

% \begin{figure*}[ht]
% \centering
% \includegraphics[width=0.5\textwidth]{figures/ccf_gauss.eps}
% \caption{Estimated cross-correlation (solid lines) on synthetic realizations of the process with $v=2w$. The cross-correlation functions $\wh{R}_x$ and $\wh{R}_y$ are estimated between the signal measured at points $P_0=(0,0)$ and $P_x=(1,0)$, and $P_0=(0,0)$ and $P_y=(0,1)$ respectively, where distances are measured in units of pulse sizes $\ellx = \elly$. Time lags $\triangle_t$ are normalized with the duration time $\taud$ defined in \Secref{sec.test}. Theoretical values following \Eqref{eq.ccf.2g} are also shown (dashed lines).}
% \label{fig.ccf}
% \end{figure*}

\section{Time delay estimation method}\label{sec.estimation}

In this section, we describe a method for velocity estimation based on the process recorded in a coarse array of measurement points.

\subsection{Aligned measurement grids}\label{sec.estimation.aligned}

\begin{figure}[tb]
\centering
\def\svgwidth{.5\columnwidth}
%% Creator: Inkscape 1.3.2 (1:1.3.2+202311252150+091e20ef0f), www.inkscape.org
%% Accompanies image file 'gpi_tex.pdf' (pdf, eps, ps)
\begingroup%
  \makeatletter%
  \providecommand\color[2][]{%
    \errmessage{(Inkscape) Color is used for the text in Inkscape, but the package 'color.sty' is not loaded}%
    \renewcommand\color[2][]{}%
  }%
  \providecommand\transparent[1]{%
    \errmessage{(Inkscape) Transparency is used (non-zero) for the text in Inkscape, but the package 'transparent.sty' is not loaded}%
    \renewcommand\transparent[1]{}%
  }%
  \providecommand\rotatebox[2]{#2}%
  \newcommand*\fsize{\dimexpr\f@size pt\relax}%
  \newcommand*\lineheight[1]{\fontsize{\fsize}{#1\fsize}\selectfont}%
  \ifx\svgwidth\undefined%
    \setlength{\unitlength}{333.16695326bp}%
    \ifx\svgscale\undefined%
      \relax%
    \else%
      \setlength{\unitlength}{\unitlength * \real{\svgscale}}%
    \fi%
  \else%
    \setlength{\unitlength}{\svgwidth}%
  \fi%
  \global\let\svgwidth\undefined%
  \global\let\svgscale\undefined%
  \makeatother%
  \begin{picture}(1,0.8121709)%
    \lineheight{1}%
    \setlength\tabcolsep{0pt}%
    \put(0,0){\includegraphics[width=\unitlength,page=1]{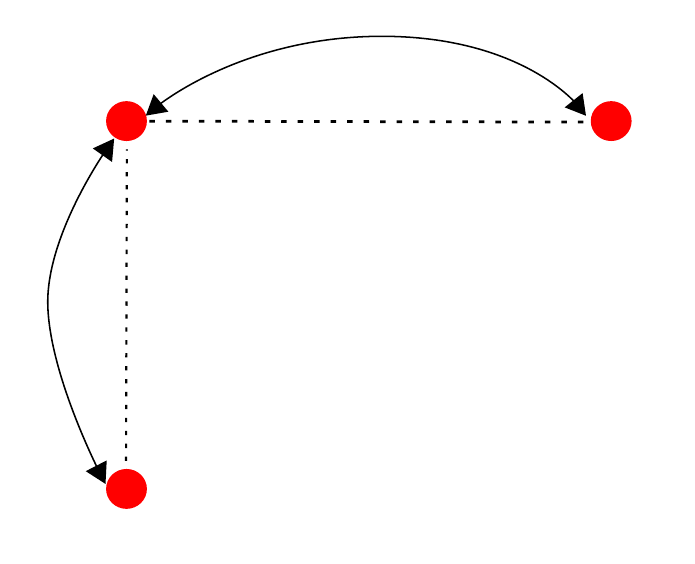}}%
    \put(0.52791926,0.66013653){\makebox(0,0)[lt]{\lineheight{1.25}\smash{\begin{tabular}[t]{l}$\Delta_x$\end{tabular}}}}%
    \put(0.142861,0.35335601){\rotatebox{90}{\makebox(0,0)[lt]{\lineheight{1.25}\smash{\begin{tabular}[t]{l}$\Delta_y$\end{tabular}}}}}%
    \put(0.16468355,0.68856985){\makebox(0,0)[lt]{\lineheight{1.25}\smash{\begin{tabular}[t]{l}$P_0$\end{tabular}}}}%
    \put(0.8740016,0.6820705){\makebox(0,0)[lt]{\lineheight{1.25}\smash{\begin{tabular}[t]{l}$P_x$\end{tabular}}}}%
    \put(0.15262226,0.01534692){\makebox(0,0)[lt]{\lineheight{1.25}\smash{\begin{tabular}[t]{l}$P_y$\end{tabular}}}}%
    \put(0.52732518,0.79144667){\makebox(0,0)[lt]{\lineheight{1.25}\smash{\begin{tabular}[t]{l}$\wh{R}_x(\tlag)$\end{tabular}}}}%
    \put(0.03065664,0.35116296){\rotatebox{90}{\makebox(0,0)[lt]{\lineheight{1.25}\smash{\begin{tabular}[t]{l}$\wh{R}_y(\tlag)$\end{tabular}}}}}%
  \end{picture}%
\endgroup%

\caption{Basic setup for three-point velocity estimation: a reference point $P_0$ and two non-aligned points $P_x$ and $P_y$.}
\label{fig.gpi}
\end{figure}
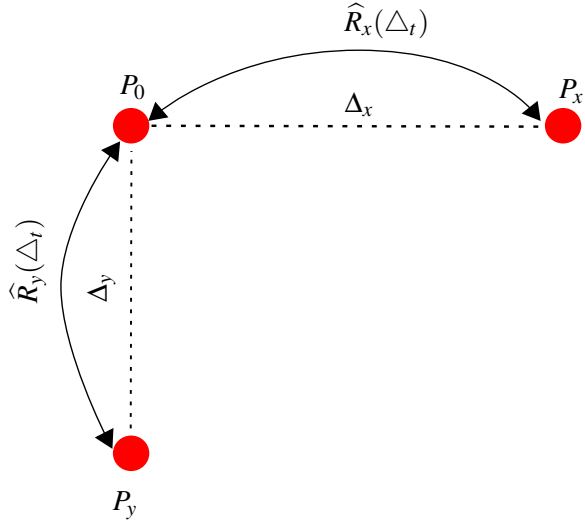

The basic setup is illustrated in \Figref{fig.gpi}. Consider three spatially separated measurement points, $P_0$, $P_x$ and $P_y$. We take $P_x$ to be separated by a horizontal distance $\dx$ from $P_0$ and $P_y$ to be separated by a vertical distance $\dy$ from $P_0$. A generalization of this to pixel grids that are not orthogonal and not aligned with the coordinate system is straightforward and provided in \Secref{sec.generalization}.

We denote the estimated cross-correlation functions between the signals measured at $P_0$ and $P_x$ and between the signals measured at $P_0$ and $P_y$ by $\wh{R}_x(\tlag)$ and $\wh{R}_y(\tlag)$, respectively. Maximization of these cross-correlation functions with respect to the time lag $\tlag$ will yield time lags $\tau_x$ and $\tau_y$, respectively. Assuming that the pulses are tilted in the direction of propagation, $\tilt=\vdir$, we demonstrate in \Appref{app.ccf} that the time lag $\taumax(\xlag, \ylag)$ that maximizes the model cross-correlation function, \Eqref{eq.ccf.2g}, is independent of the sizes $\ellx$ and $\elly$,
\begin{equation}\label{eq.tmax}
    \taumax(\xlag, \ylag) = \frac{v \xlag + w \ylag }{v^2 + w^2}.
\end{equation}
In particular, for $\ylag=0$ the cross-correlation function is maximum for time lag ${\tau}_x=v\xlag/(v^2+w^2)$, while for $\xlag=0$ the cross-correlation function is maximum for time lag ${\tau}_y=w\ylag/(v^2+w^2)$. To give the velocity component estimates, these relations can be inverted with respect to $v$ and $w$,
\begin{subequations}\label{eq.vw.simpl}
\begin{equation}
    \wh{v} = \frac{\dx / \tau_x}{1 + \left(\frac{\dx / \tau_x}{\dy / \tau_y} \right)^2},  
\end{equation}
\begin{equation}
    \wh{w} = \frac{\dy / \tau_y}{1 + \left(\frac{\dy / \tau_y}{\dx / \tau_x} \right)^2}.
\end{equation}
\end{subequations}
In \Secref{sec.test}, we will apply these relations to estimate the velocity components from realizations of the process to assess the validity and limits of the method. In the case $\alpha\neq\vdir$, an error will be caused by the barberpole effect if $\ellx/\elly\neq 1$. We study the extent of these errors in \Secsref{sec.elipse} and \ref{sec.tilt}.

% TODO: Set \tau_{\text{max}} in R to get an estimate of how big can ell/Delta be so that R is bigger than 0.5.
\subsection{Non-aligned measurement grids}\label{sec.generalization}

\begin{figure}[ht]
\centering
\def\svgwidth{.5\columnwidth}
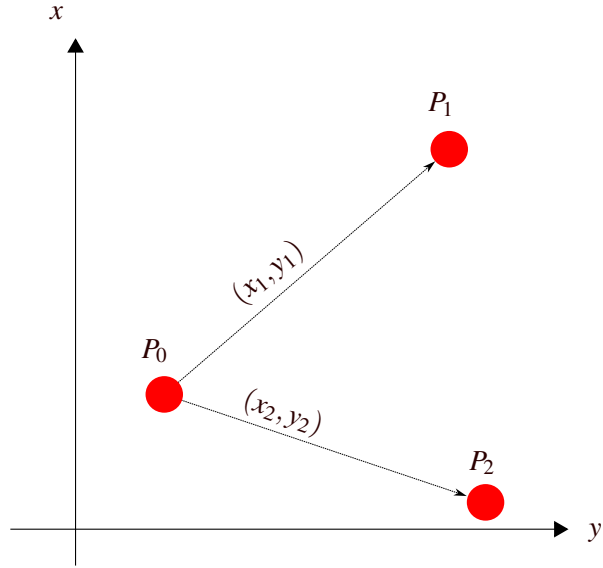
\caption{Basic setup for three-point velocity estimation where the three measurement points are not aligned with the coordinate axes.}
\label{fig.grid}
\end{figure}

The above treatment is simplified by assuming that the measurement points are aligned with the coordinate axes, that is, $P_x$ and $P_y$ are separated from $P_0$ horizontally and vertically, respectively. We present the results this way since that is the case for many diagnostics and results in simpler expressions for the estimates. The generalization to non-orthogonal measurement grids, such as depicted in \Figref{fig.grid}, is straightforward. Equation (\ref{eq.tmax}) can be evaluated for any two given pairs of points and the time lags for maximum cross-correlation can be inverted with respect to the velocity components leading to the velocity estimates
\begin{subequations}
\begin{equation}
    \wh{v} = \frac{(\tau_2 y_1 - \tau_1 y_2)(x_2 y_1 - x_1 y_2)}{(\tau_2 x_1 - \tau_1 x_2)^2 + (\tau_2 y_1 - \tau_1 y_2)^2 }, \\
\end{equation}
\begin{equation}
    \wh{w} = \frac{(\tau_2 x_1 - \tau_1 x_2)(x_1 y_2 - x_2 y_1)}{(\tau_2 x_1 - \tau_1 x_2)^2 + (\tau_2 y_1 - \tau_1 y_2)^2}.
\end{equation}
\end{subequations}
The numerator appearing in both components expresses the requirement that the measurement points should not be aligned, ($x_1y_2 - x_2y_1 \neq 0$), as otherwise, the time delays $\tau_1$ and $\tau_2$ would provide redundant information.

\section{Application on synthetic data}\label{sec.test}
\newcommand{\tf}[1]{\textfrak{#1}}

%\subsection{Model normalization}

We normalize the model in terms of a characteristic length $\ell = \sqrt{\ellx \elly}$, and a characteristic time $\taud = \ell/u$, where $u^2 = v^2 + w^2$. The amplitudes are normalized with the average amplitude $\ave{a}$. With this convention, the process defined by \Eqref{eq.2d} can be written as
\begin{equation}
    \Phi_K(x, y, t) = \sum_{k=1}^{K} a_k \varphi\left(\frac{x - v(t - \tk)}{\ellx}, \frac{y - \yk - w(t - \tk)}{\elly} \right)
\end{equation}
where all variables are normalized with their respective characteristic values and are thus dimensionless. We employ the same symbols ($x$, $y$, $t$, $a_k$, $v$, $w$, $\ellx$, $\elly$, $\tk$ and $\yk$) as in the non-normalized process for simplicity of notation. All numerical simulations of the process are performed using these dimensionless variables.

\begin{center}
\begin{table}[h!]
\begin{tabular}{ |c |c| c |}
\hline
 Signal duration & $T/\taud$ & $10^3$ \\
 Number of pulses & $K$ & $10^3$ \\ 
 Aspect ratio & $\ellx/\elly$ & $1$ \\  
 Velocity angle & $\vdir$ & $0, \pi/6, -\pi/3$    \\
 Sampling time & $\dt / \taud$ & $10^{-2}$ \\
 Spatial resolution & $\Delta / \ell$ & $1$ \\
 Domain size & $L / \ell$ & $10$ \\
\hline
\end{tabular}
\caption{Input parameters for realizations of the stochastic process together with their default values in dimensional units.}
\label{table.input}
\end{table}
\end{center}

In all cases, the amplitudes are taken to be the same for all pulses, in dimensional units $a_k=\ave{a}$. The initial vertical positions $\yk$ at time $t=\tk$ are uniformly distributed on the interval $[0, L]$, with $L/\ell = 10$. Except for in \Secref{sec.tilt}, the pulses are by default tilted in their direction of propagation $\alpha=\vdir$, as illustrated in \Figref{fig.parameters}.

Realizations of this model are performed with the resulting signal measured at three points, $P_0$, $P_x$ and $P_y$, separated by a distance $\Delta_x = \Delta_y = \Delta$ as shown in \Figref{fig.gpi}. The relevant input parameters of the realizations, together with their default values, are given in Table \ref{table.input}.

In order to obtain good estimates of the performance of the method, ten independent simulations of the process are performed for each combination of parameters. For each realization labelled $i$, the velocity components, $\wh{v}_i$ and $\wh{w}_i$, are estimated with the three-point time delay method, and the mean-square error ($\MSE$) is computed as
\begin{equation}\label{eq.mse}
    \MSE = \frac{1}{N} \sum_{i=1}^N \frac{(\wh{v}_i-v)^2+(\wh{w}_i-w)^2}{u^2}.
\end{equation}
In order to not confound precision with accuracy we are also interested in the standard deviation of the estimates, and so we introduce
\begin{equation}
    \sigma_\MSE^2 = \frac{1}{N} \sum_{i=1}^N \frac{(\wh{v}_i-\ave{\wh{v}})^2 + (\wh{w}_i-\ave{\wh{w}})^2}{u^2} ,
\end{equation}
where $\ave{\wh{v}}$ and $\ave{\wh{w}}$ are the averaged estimated velocity components. We will consider the method to be reliable in the cases that the mean square error is lower than the arbitrary threshold $\MSE < 10^{-1}$. This is done in order to provide bounds on the parameter space where the method is considered reliable. Next, if $\MSE$ is high while $\sigma_\MSE$ is low, we conclude that the estimation is biased. In each of the following subsections, we will vary different input parameters and present the resulting error and standard deviation of the estimation method. The results will be presented for three different velocity directions as presented in Table \ref{table.input}, with $v/u = 1$, $v/u=1/2$ and $w/u=-1/2$, respectively (these correspond to $\vdir = 0$, $\pi/6$ and $-\pi/3$, respectively). The estimation method is implemented to not output any estimates if the cross-correlation maxima are below a user-defined threshold or if both time delays, $\tau_x$ and $\tau_y$, are smaller than the sampling time $\dt$. In the following studies and in order to illustrate the performance of the method, we relax these safeguards while indicating where they are not fulfilled: this is an important aspect of the method, showing that many of the cases where the method fails can be anticipated a priori.

\subsection{Signal duration}

\begin{figure*}[tb]
\centering
\includegraphics[width=\textwidth]{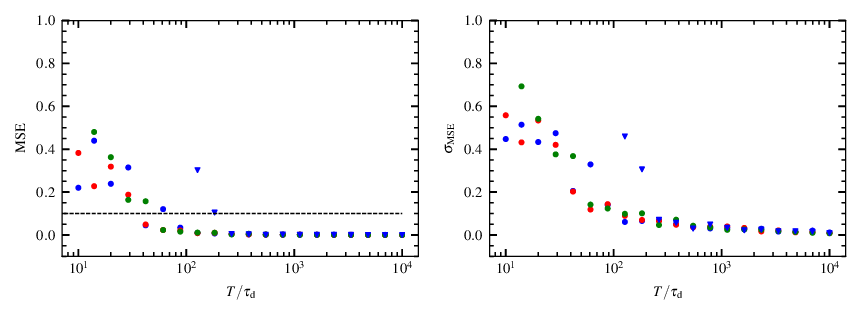}
\caption{Left: Mean square error of the three-point velocity estimation for realizations of the process with varying signal duration and a fixed ratio of signal duration to number of pulses $T/K$. Each plot symbol represents the average mean-square error of 10 independent realizations. Blue, red and green symbols represent realizations with $\vdir = 0$, $\pi/6$ and $-\pi/3$, respectively. The blue triangles represent realizations with $\vdir = 0$ and a domain size $L/\ell=10^2$. Right: Mean square error standard deviation over the $10$ simulations.}
\label{fig.signal_length}
\end{figure*}

We start by considering the effect of the signal duration $T$ while keeping the average pulse density $T/K=\taud$ fixed. The results for the mean-squared error are shown in \Figref{fig.signal_length}. The method is unbiased and it is reliable for signal durations roughly $T/\taud>10^2$. This can be interpreted as follows: enough pulses, $\sim 10$, need to be detected by the measurement points in order to accurately estimate the pulse velocities. Since the simulation domain in dimensional units is $L/\ell=10$, we need $K > 10 L/\ell = 10^2$. Since in our simulations we keep $T/K=\taud$, this translates into $T/\taud > 10^2$, which is consistent with the plotted results. In order to verify this interpretation, we made a series of simulations with $L/\ell=10^2$, expecting that the signal duration required for a reliable estimate to be longer. This is indeed the case as shown by the downward-pointing triangles in \Figref{fig.signal_length}.

\subsection{Number of pulses}

\begin{figure*}[tb]
\centering
\includegraphics[width=\textwidth]{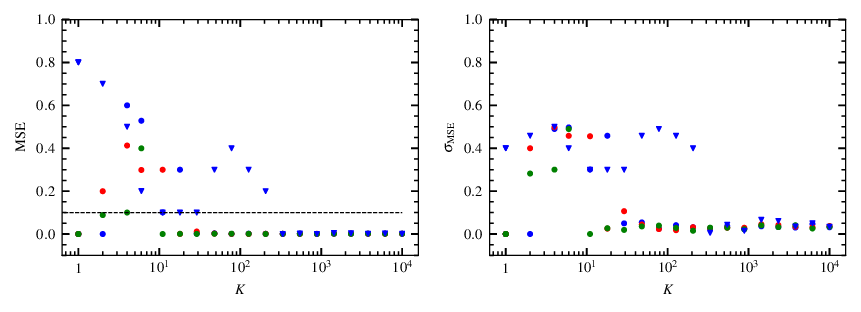}
\caption{Left: Mean square error of the three-point velocity estimation for realizations of the process with varying number of pulses for a fixed signal duration. Each plot symbol represents the average mean-square error of 10 independent realizations. Blue, red and green symbols represent realizations with $\vdir = 0$, $\pi/6$ and $-\pi/3$, respectively. The blue triangles represent realizations with $\vdir = 0$ and a domain size $L/\ell=10^2$. Right: Mean square error standard deviation over the $10$ simulations.}
\label{fig.number_pulses}
\end{figure*}

Consider now the effect of varying the number of pulses $K$ while keeping the signal duration $T$ fixed. The results are presented in \Figref{fig.number_pulses} and show that the method is reliable for $K>10^2$. As discussed above, this is likely due to the choice of the spatial extent of the simulation domain, $L/\ell = 10$, which renders $K=10^2$ as the threshold under which not enough pulses are detected by the measurement points for a reliable velocity estimate. We verify this interpretation by making a series of simulations with $L/\ell=10^2$. As shown by the triangles in \Figref{fig.number_pulses}, the number of pulses the needs to be larger in order for the method to produce accurate velocity estimates.

\subsection{Spatial resolution}

\begin{figure*}[tb]
\centering
\includegraphics[width=\textwidth]{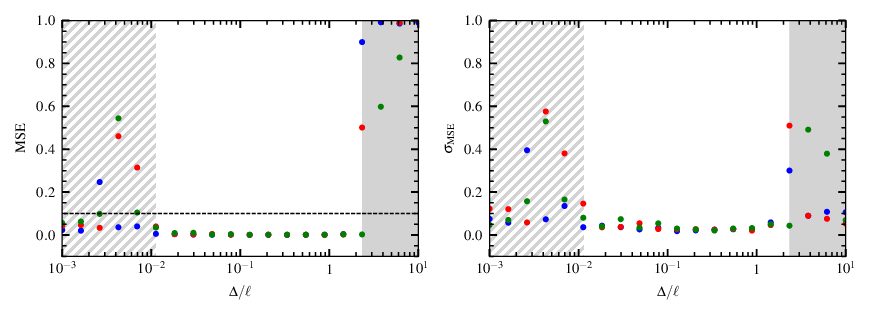}
\caption{Left: Mean square error of the three-point velocity estimation for realizations of the process with different spatial resolutions $\Delta/\ell$. Each plot symbol represents the average mean-square error of 10 independent realizations. Blue, red and green symbols represent realizations with $\vdir = 0$, $\pi/6$ and $-\pi/3$, respectively. The gray-shaded region to the right represents the parameter space where the cross-correlation maxima fall under $0.25$. The hatched region to the left represents the parameter space where the maximum estimated time delay is shorter than the sampling time. Right: Mean square error standard deviation over the $10$ simulations.}
\label{fig.delta}
\end{figure*}

In this subsection, we investigate the role of spatial resolution, that is, the distance $\Delta$ between measurement points. We made realizations of the model where the distance $\Delta$ is varied. The results are presented in \Figref{fig.delta}, which show that the method is reliable for approximately $0.02 < \Delta/\ell < 2$. For small values of $\Delta$, the results show high uncertainties in the estimation. The cause of these higher uncertainties is that the estimated time delays $\tau_x$ and $\tau_y$ are of the same order of magnitude as the sampling time $\dt$, which leads to high uncertainties in the estimations of $\tau_x$ and $\tau_y$. The grey hatched region in \Figref{fig.delta} represents the area where the maximum estimated time delay, $\max(\tau_x, \tau_y)$, is smaller than the sampling time $\dt$. The transit time of the pulses between measurement points is given by \Eqref{eq.tmax}. Assuming $v \sim w \sim u$ we can write $\tau_x \sim \tau_y \sim \Delta/u$. The requirement that the transit time is larger than the sampling time, $\tau > \dt$, is equivalent to $\Delta / \ell > \dt / \taud = 10^{-2}$. This is consistent with the results in \Figref{fig.delta}. In practice, for higher-resolution imaging data, this potential pitfall can be easily solved by employing measurement points that are not nearest neighbours from each other, but that lie at a distance such that the cross-correlation function between their signals is maximized for a time delay larger than the sampling time.

For large values of $\Delta/\ell$, the signals recorded as different points become uncorrelated, as pulses detected at a given point may not be detected at another. This can be easily anticipated by the safeguard of imposing a lower threshold on the cross-correlation maxima under which no velocity estimation should be performed. For $\Delta/\ell > 2$ we obtain cross-correlation maxima well under $0.25$. For experimental data where other sources of correlation might be present, a higher threshold can be advised, typically $0.5$. In conclusion, the requirement $0.02 < \Delta/\ell < 2$ can easily be checked on empirical data by setting a minimum threshold on the times maximizing the cross-correlation functions, $\tau_x$ and $\tau_y$, and on the maximum value of the cross-correlation functions.

\subsection{Temporal resolution}

\begin{figure*}[tb]
\centering
\includegraphics[width=\textwidth]{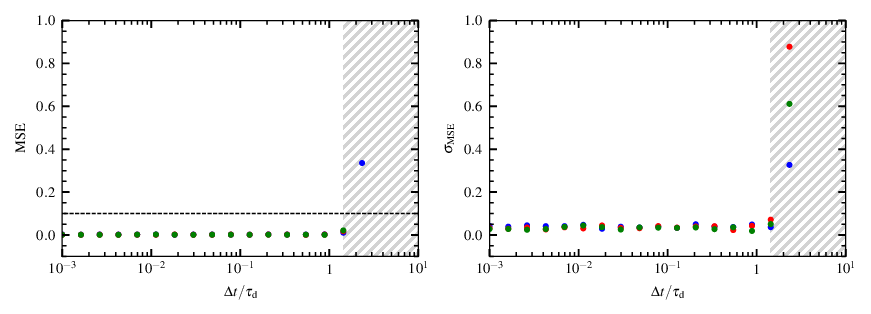}
\caption{Left: Mean square error of the three-point velocity estimation for different realizations of the process with different values of the sampling time $\dt/\taud$. Each plot symbol represents the averaged mean-square error of 10 independent realizations. Blue, red and green represent realizations with $\vdir = 0$, $\pi/6$ and $-\pi/3$, respectively, where $v$ and $w$ are the average velocity components. The hatched region represents the parameter space where the maximum estimated time delay is lower than the sampling time. Right: Mean square error standard deviation over the $10$ simulations.}
\label{fig.delta_t}
\end{figure*}

Turning now to the effect of the temporal resolution, we perform realizations with different values of the normalized sampling time $\dt/\taud$. The results are presented in \Figref{fig.delta_t}. From this, it follows that the method is reliable up to sampling times $\dt \sim \taud$, at which point the cross-correlation function is maximized at time lags of the order of the sampling time, leading to high uncertainties in the estimation of the time lags and the resulting velocities. For $\dt/\taud \gg 1$ the errors become very large and lie outside the plot domain presented in \Figref{fig.delta_t}. We conclude that the method is reliable when the underlying velocities lead to transit times $\taud$ larger than the sampling time $\dt$, and this can be guaranteed by imposing the safeguard $\max(\tau_x, \tau_y)>\dt$ as discussed above and indicated by the hatched region in \Figref{fig.delta_t}.

\subsection{Random velocities}

\begin{figure*}[tb]
\centering
\includegraphics[width=\textwidth]{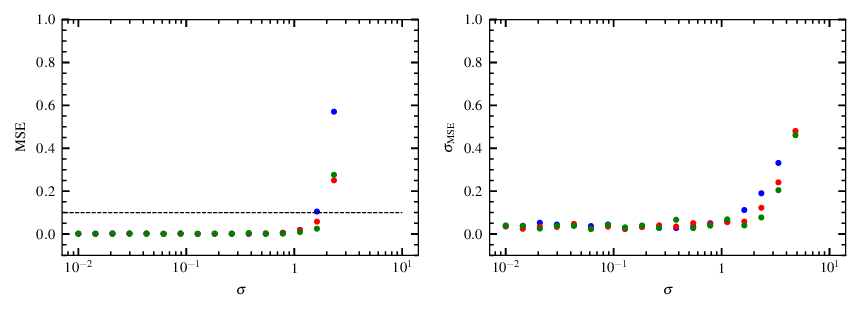}
\caption{Left: Mean square error of the three-point velocity estimation for different realizations of the process with uniformly distributed velocity components for different values of the width parameter $\sigma$. Each plot symbol represents the averaged mean-square error of 10 independent realizations. Blue, red and green represent realizations with $\vdir = 0$, $\pi/6$ and $-\pi/3$, respectively, where $v$ and $w$ are the average velocity components. Right: Mean square error standard deviation over the $10$ simulations.}
\label{fig.realization_rand_unif}
\end{figure*}

In this subsection, we consider the effect of a random distribution of pulse velocities given by
\begin{subequations}\label{eq.random_velocities}
\begin{gather}
    v_k = \ave{v} + u\tilde{v}_k ,
    \\
    w_k = \ave{w} + u\tilde{w}_k ,
\end{gather}
\end{subequations}
where each average velocity component has an added, normalized zero-mean uniformly distributed random deviation with width parameter $\sigma>0$,
\begin{equation}
    P_{\tilde{v}}(\tilde{v};\sigma) = \begin{cases}
        1/2\sigma\, & \text{if } \lvert{\tilde{v}}\rvert \leq \sigma , \\
        0\, &  \text{if } \lvert{\tilde{v}}\rvert > \sigma ,
    \end{cases}
\end{equation}
and similar for $\tilde{w}$. In \Eqref{eq.random_velocities} we have defined $u^2=\ave{v}^2+\ave{w}^2$.
%$\tilde{v}_k, \tilde{w}_k \sim U([-\sigma/2, +\sigma/2])$, where the uniform probability distribution function is given by
%\begin{equation}
%    f(x) = \begin{cases}
%        \frac{1}{\sigma}\, & \text{if } -\frac{\sigma}{2} < x < \frac{\sigma}{2} \\
%        0\, & \text{otherwise}
%    \end{cases}
%\end{equation}
%and where $0<\sigma<2$ is the width parameter of the uniform distribution. 
The results from the velocity estimation are presented in \Figref{fig.realization_rand_unif}. The method remains unbiased and is reliable up to $\sigma \approx 2$, for which the broadness of the velocity distribution is larger than the average velocity. The cross-correlation maxima remains roughly unchanged for different values of $\sigma$. We conclude that the three-point TDE method works well to estimate the average velocity even for a broad distribution of pulse velocities up to broadness comparable to the average value.

\subsection{Additive noise}

\begin{figure*}[tb]
\centering
\includegraphics[width=\textwidth]{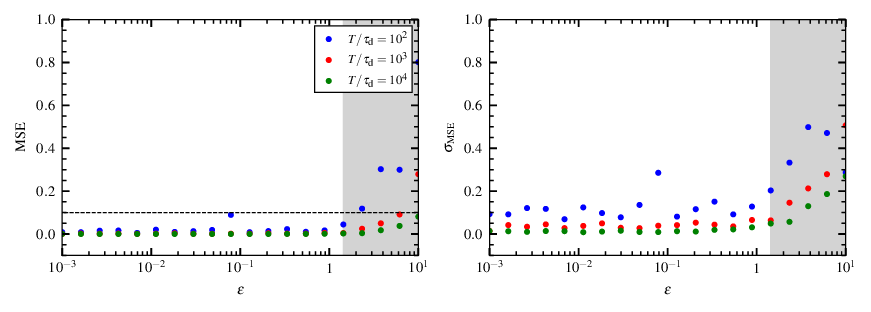}
\caption{Left: Mean square error of the three-point velocity estimation for realizations of the process with added spatially and temporally uncorrelated white noise for different values of the noise-to-signal ratio $\epsilon$. Each plot symbol represents the averaged mean-square error of 10 independent realizations. Blue, red and green symbols represent realizations with different signal durations as indicated in the legend. The number of pulses to signal duration ratio, $K/T$, is kept constant. The velocity direction is kept constant at $\vdir=0$. The gray-shaded region represents the parameter space where the cross-correlation maxima fall under $0.25$. Right: Mean square error standard deviation over the $10$ simulations.}
\label{fig.noise}
\end{figure*}

We next consider the effect of additive noise to the simulation data, a situation commonly encountered in experimental measurements \cite{theodorsen_statistical_2017,garcia_sol_2017}. To this end, we add spatially and temporarily uncorrelated white noise to each measurement point with a noise-to-signal ratio given by a parameter $\epsilon$. The results of the velocity estimation are presented in \Figref{fig.noise} for different signal durations. The method is reliable up to noise-to-signal ratios of $\epsilon \sim 5$, at which point the cross-correlation maxima are well below $0.25$. Longer signal durations with $T/\taud = 10^5$ (green) lead to marginally better estimates than shorter signal durations with $T/\taud = 10^2$ (blue).

\subsection{Correlated amplitudes and velocities}

Since the presented method is based on the cross-correlation between signals measured at different points, we expect pulses with higher amplitude to contribute more to the resulting estimated cross-correlation function. Thus, in the case that amplitudes are correlated with velocities, we expect the method to be biased towards the velocities of the higher-amplitude pulses. In this subsection, we present realizations of the process in which the amplitudes of the pulses are uniformly distributed with mean values $\aave$ and width parameter $\sigma$,
%on an interval
\begin{equation}
    \aave P_{\tilde{a}}(\tilde{a};\aave,\sigma) = \begin{cases}
        1/2\sigma\, & \text{if } 1-\sigma \leq a/\aave \leq 1+\sigma , \\
        %\lvert{\tilde{v}}\rvert \leq \sigma , \\
        0\, &  \text{otherwise} . % \lvert{\tilde{v}}\rvert > \sigma ,
    \end{cases}
\end{equation}
%\begin{equation}
%    a_k \sim U([\ave{a}(1-\sigma/2), \ave{a}(1+\sigma/2)]),
%\end{equation}
%where $\sigma$ is the width parameter. 
The velocities are taken to be proportional to the amplitudes
\begin{subequations}
\begin{gather}
    v_k/\ave{v} = a_k/\ave{a}.
    \\
    w_k/\ave{w} = a_k/\ave{a}.
\end{gather}
\end{subequations}
For smaller values of $\sigma$, the distributions are narrow and we do not expect the results to differ from the base case study. For large values of $\sigma$, the amplitude and velocity distribution become broad and we expect the method to be biased towards the velocities of high-amplitude pulses.

\begin{figure*}[tb]
\centering
\includegraphics[width=\textwidth]{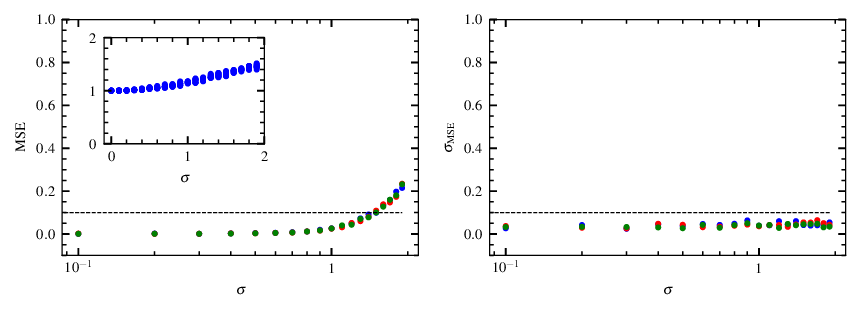}
\caption{Left: Mean square error of the three-point velocity estimation for realizations of the process with the pulse velocities proportional to the amplitudes for different values of the distribution width parameter $\sigma$. Each plot symbol represents the average mean-square error of 10 independent realizations. Blue, red and green symbols represent realizations with $\vdir = \arctan \ave{w}/\ave{v} = 0$, $\pi/6$ and $-\pi/3$, respectively, for the average velocity components. Inset left: Estimated horizontal velocity components $\wh{v}_i$ for $\vdir=0$. Right: Mean square error standard deviation over the $10$ simulations.}
\label{fig.corr}
\end{figure*}

The results from the velocity estimation are presented in \Figref{fig.corr} and agrees with this interpretation, the method is reliable up to unity values of $\sigma$, where the broadness of the correlation is of the order of the average value and the method is biased towards high-amplitude and high-velocity pulses, thus overestimating the average velocity.

\subsection{Elongated pulses}\label{sec.elipse}

\begin{figure*}[tb]
\centering
\includegraphics[width=\textwidth]{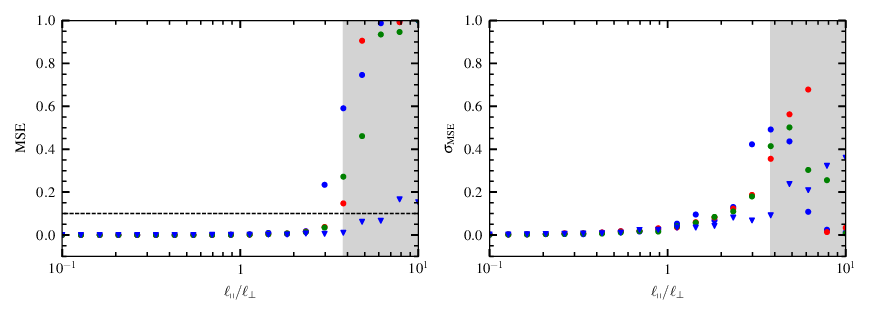}
\caption{Left: Mean square error of the three-point velocity estimation for realizations of the process with different values of the pulse aspect ratio $\ellx/\elly$. Each plot symbol represents the average mean-square error of 10 independent realizations. Blue, red and green symbols represent realizations with $\vdir = 0$, $\pi/6$ and $-\pi/3$, respectively. Blue triangles represent a series of simulations with enhanced spatial resolution, $\Delta/\ell = 1/5$. The grey-shaded region to the right represents the parameter space where the cross-correlation maxima falls under $0.25$ (it does not apply to the case with enhanced spatial resolution). Right: Mean square error standard deviation over the $10$ simulations.}
\label{fig.elongated_aligned}
\end{figure*}

In this subsection, we consider the effect of elongated pulses with $\ellx/\elly \neq 1$. For all the realizations in this subsection, the pulses are taken to be aligned with their direction of motion, $\alpha=\beta$. Thus, $\elly$ describes the spatial extent to which the pulse can be measured, as pulses with small or large $\elly$ will be respectively less or more likely to be detected by a set of measurement points. Lower values of $\elly$ will thus lead to less correlated signals. The results from the velocity estimation are presented in \Figref{fig.elongated_aligned}, where we include a series of realizations with increased spatial resolution, $\Delta/\ell = 1/5$, marked by triangle symbols. For the base case $\Delta = \ell$, the error is larger for large values of $\ellx/\elly$. This is explained by a drop in the cross-correlation value between the different measurement signals: for smaller values of $\elly$, the pulse is detected in fewer measurement points and the signals become uncorrelated. However, the series of realizations with improved spatial resolution, indicated by the downward-pointing triangles, barely improves the error even though the cross-correlation maxima are substantially higher (higher than $0.8$ for all the cases considered). One possible reason is that the characteristic size, $\ell = \sqrt{\ellx \elly}$, used to normalize space, is kept constant. As a result, a lower $\elly$ implies a higher $\ellx$, which flattens the cross-correlation function near the maxima and increases the uncertainty in estimating the time lag for the maximum.

\subsection{Tilted pulses}\label{sec.tilt}

\begin{figure*}[tb]
\centering
\includegraphics[width=\textwidth]{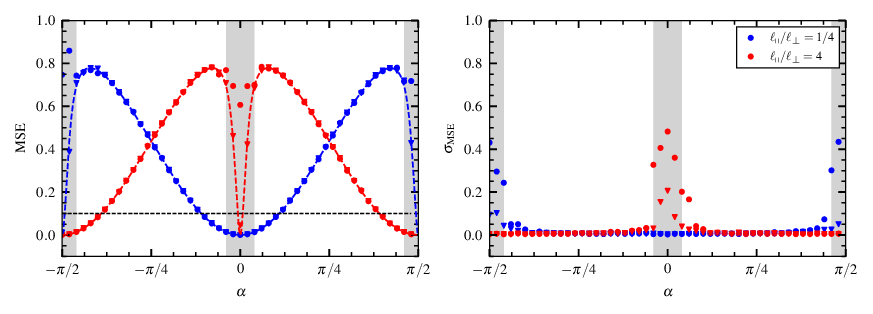}
\caption{Left: Mean square error of the three-point velocity estimation for realizations of the process with different values of the tilt angle $\tilt$. Each plot symbol represents the average mean-square error of 10 independent realizations. All realizations are performed with  $\vdir = 0$. Blue and red symbols represent realizations with $\ellx/\elly = 1/4$ and $4$, respectively. Blue and red triangles represent simulations with enhanced spatial resolution, $\Delta/\ell = 0.2$, which lead to the same errors in most cases. The gray-shaded region represents the parameter space where the cross-correlation maxima fall under $0.25$ for the case of $\ellx/\elly=4$ near $\tilt=0$, and for the case of $\ellx/\elly=1/4$ near $\tilt=\pm\pi/2$. The simulations with enhanced spatial resolution have high cross-correlation values in all the explored parameter regimes. Dashed lines represent the analytical expression for the time delays derived in \Appref{app.ccf}. Right: Mean square error standard deviation over the $10$ simulations.}
\label{fig.tilt}
\end{figure*}

Finally, we consider the effect of tilted pulses, varying the angle $\tilt$ between the pulse axis and the horizontal axis, as shown in \Figref{fig.parameters}. We made realizations of the process with horizontally propagating pulses, $\vdir=0$, thus $w=0$, and $\tilt \in [-\pi/2, \pi/2]$. The resulting errors of the velocity estimation method are presented in \Figref{fig.tilt} for the cases $\ellx/\elly = 4$ and $1/4$. Other velocity directions give identical results. The results show that the method is reliable only for $\tilt-\vdir\approx 0$ or $\tilt-\vdir\approx \pm \pi/2$. This is due to the so-called barberpole effect. For small values of $\tilt-\vdir$, the bias is strongly dependent on the aspect ratio $\ellx/\elly$. For $\ellx/\elly = 1/4$, the method is reliable ($\MSE<0.1$) for approximately $|\tilt-\vdir|<\pi/8$, whereas for $\ellx/\elly=4$ the method rapidly becomes unreliable for $\tilt-\vdir\neq 0$. As expected, the case of $\ellx/\elly = 1/4$ is equivalent to $\ellx/\elly=4$ for $\tilt-\vdir=\pm\pi/2$, as that effectively interchanges the parallel and perpendicular axes. In all cases, the mean-square error is less than unity.

\section{Discussion and conclusions}\label{sec.dc}

The transport of particles and heat in the scrape-off layer of magnetically confined plasmas is dominated by the propagation of coherent pulses of enhanced plasma density. We have investigated a model describing the fluctuations resulting from a superposition of pulses propagating in two dimensions. This model is an extension of models extensively studied in zero- and one-space dimensions \cite{garcia_stochastic_2012,kube_convergence_2015,garcia_stochastic_2016,garcia_auto-correlation_2017,theodorsen_statistical_2017,militello_scrape_2016,militello_relation_2016,theodorsen_level_2018,theodorsen_probability_2018,ahmed_reconstruction_2023,losada_stochastic_2023,losada_stochastic_2024}.

In the case of Gaussian pulses, a degenerate distribution of pulse velocities, and uncorrelated pulse amplitudes and velocities, an expression for the cross-correlation function can be obtained analytically. From this expression, it is straightforward to compute the time $\taumax(\xlag, \ylag)$ that maximizes the cross-correlation function. This time depends on the underlying two-dimensional velocity field. Thus, two independent estimations of $\taumax(\xlag, \ylag)$ can determine the velocity components. A minimum of three non-aligned measurement points is required in order to obtain two independent estimations of $\taumax(\xlag, \ylag)$, and so, the proposed method is referred to as a three-point velocity estimation method. Given the low number of measurement points required for the estimation, this method is particularly useful in the case of coarse-grained imaging diagnostics, such as avalanche photodiode gas-puff imaging \cite{cziegler_experimental_2010,zweben_invited_2017,offeddu_gas_2022,terry_realization_2024} or beam emission spectroscopy \cite{mckee_beam_1999,wang_beam_2023}, where more robust velocity estimation methods do not apply due to the low spatial resolution \cite{sierchio_comparison_2016}.

We studied the performance of the method on synthetic data for a wide range of scenarios. Firstly, our findings show that the signal duration should be at least long enough to contain a significant amount of pulses. The method works for any degree of pulse overlap. Next, we find that the separation between measurement points should be at least the distance travelled by the pulses during a sampling time regardless of the direction of motion. In practical cases, this can be verified by ensuring that the times maximizing the cross-correlation functions are larger than the sampling times. Additionally, we find that the distance between measurement points must be small enough to ensure the correlation is high enough to estimate the maximum of the cross-correlation function reliably. In the case of Gaussian pulses, this implies a separation shorter than approximately two pulse sizes. This can be verified by assuring that the value of the cross-correlation maxima is sufficiently high. In the case of synthetic data, this value is set at $0.25$, but higher values may be required for experimental data. Similarly, we find that the temporal resolution should be smaller than the time the pulses spend to travel between measurement points, $\dt > \taud$. We find that the method is also robust when the pulse velocities are randomly distributed, providing the correct mean velocity values even in cases where the broadness of the velocity distribution is of order unity. Next, we find that the method is robust against additive noise, providing correct estimates up to noise-to-signal ratios of approximately $5$. Additionally, we have investigated the case where the pulses have a temporal amplitude decay modeled as a linear damping \cite{losada_stochastic_2023}. We found that the method is robust against temporal variations of the pulse amplitudes leading to no effect in the velocity estimation. Lastly, a distribution of pulse sizes has also been investigating and has no effect in the velocity estimation results.

Since the method is based on time-delay estimation from cross-correlation functions, we expect the velocities of larger-amplitude events to contribute more to the resulting velocity estimates, as large-amplitude pulses have a larger contribution to the cross-correlation function. We find that this is indeed the case when velocities are correlated with amplitudes, as shown by \Figref{fig.corr}, but only relevant for broadly distributed velocities. A case with exponentially distributed amplitudes and velocities $v_k/\ave{v} = a_k/\ave{a}$ and $w_k=0$ was also considered leading to large errors of up to $\MSE \sim 4$. This is due to the contribution of arbitrarily high-velocity pulses with a corresponding high amplitude. We note that theoretical blob scaling regimes often show a saturation of the velocity scaling with the amplitudes, rendering such very high-velocity blobs, nonphysical \cite{garcia_mechanism_2005,garcia_interchange_2006,kube_velocity_2011,easy_three_2014,wiesenberger_radial_2014,held_influence_2016,olsen_temperature_2016,walkden_dynamics_2016,kube_amplitude_2016,wiesenberger_unified_2017}.

When pulses are elongated, $\ellx/\elly\neq 1$, and tilted $\tilt\neq\vdir$, the so-called barberpole effect can arise \cite{fedorczak_physical_2012}. In the cases that $\ellx/\elly \gg 1$ we find a large unbiased error. This error can be understood by the fact that small values of $\elly$ make the signals recorded at different measurement points less correlated. In addition, large values of $\ellx$ lead to flatter cross-correlation maxima, making accurate estimation of such maxima more prone to errors. If the pulses do not propagate in the direction perpendicular to one of their axes, then the barberpole effect becomes the main source of error. In the worst case, this can lead up to almost unity biased $\MSE$, as shown in \Figref{fig.tilt}. The onset of the error, that is, the maximum tilting angle that renders the method still reliable depends strongly on the aspect ratio $\ellx/\elly$. For $\ellx/\elly=1/4$ the method is robust against tilting angles, remaining reliable ($\MSE < 0.1$), up to tilting angles $\theta \approx \pi/8$. On the other hand, for $\elly/\ellx=4$, small tilt angles lead to large errors. In \Appref{app.2pbp} we examine the deviations of the estimates obtained with the two-point velocity estimation method in the presence of a barberpole effect. A general comparison between the two- and three-point methods was already studied in the absence of the barberpole effect \cite{losada_three-point_2024}. Our findings show a possible use case for the two-point method: if prior knowledge of the system ensures that the structures move parallel to the separation between the two measurement points, applying the two-point method to estimate the velocity component in that direction will yield accurate results and remain unaffected by barberpole effects.

The results presented, indicating that the method is more robust for cases when the ratio $\ellx/\elly < 1$, should be contrasted with theoretical and numerical studies of blob dynamics. Blob theory suggests that as these coherent structures propagate through the SOL, they undergo a deformation process known as front steepening. This phenomenon leads to a reduction of spatial scales along the direction of propagation while simultaneously elongating the blob in the perpendicular direction \cite{garcia_interchange_2006,garcia_mechanism_2005,kube_velocity_2011,bian_blobs_2003,aydemir_convective_2005,madsen_influence_2011}. Therefore, unfavourable elongations $\ellx/\elly > 1$ may not be a concern in the case of blob propagation in the SOL.

The parameter studies conducted in this work highlight the prospects and limitations of the estimation method across different model parameter ranges. In practical applications, it is often unclear whether the data meets the method's requirements a priori. A common check involves setting a minimum threshold for the cross-correlation maximum, below which no estimation is performed. In all considered cases, we have indicated a grey-shaded region where the synthetic data fails to meet this requirement with a minimum threshold for the cross-correlation maximum of $0.25$. Our findings indicate that this requirement ensures the data meets model parameter needs in scenarios with low spatial resolution, high noise-to-signal ratio, and large $\ellx/\elly$ values. However, this requirement does not account for other model parameter deviations. Therefore, we propose an additional check based on the maximum time-delay $\max(\tau_x, \tau_y)$ used in the velocity estimation method. We require this maximum to be larger than the sampling time $\dt$. We show that, in the cases of high spatial resolution and low temporal resolution, this requirement ensures that the data fulfills the right model parameter requirements.

Thus, we implement the following safeguards: (i) a minimum threshold value for the cross-correlation function maxima, (ii) a minimum threshold for $\max(\tau_x, \tau_y)$ related to the sampling time, and (iii) that the signal duration is sufficiently long to contain a significant amount of pulses ($>10$). We have here demonstrated that these criteria ensure that the data falls within the appropriate parameter space for a wide range of model parameters. However, there are still scenarios where these two requirements are insufficient to guarantee the fulfillment of the model parameter conditions, such as broadness in the velocity distribution, pulse tilting and correlations between amplitudes and velocities.

In conclusion, the three-point velocity estimation method offers a robust and efficient approach for determining velocity components of a superposition of pulses propagating in two dimensions, even when faced with challenging scenarios such as low spatial resolution, noise, and broad velocity distributions. Our study highlights the key limitations of the method, particularly the effects of elongated and tilted pulses, where the barberpole effect can introduce significant errors. By applying minimum thresholds on the cross-correlation maxima and time-delay estimates, we ensure that the method remains reliable across a wide range of parameter spaces. However, special care must be taken in cases with strong pulse elongation or tilting. In other contributions, we will demonstrate the successful application of this improved time delay estimation method on experimental measurement data.

\section*{Acknowledgements}

This work was supported by the UiT Aurora Centre Program, UiT The Arctic University of Norway (2020).

\appendix

\section{Cross-correlation function}\label{app.ccf}

In this appendix, we will derive an analytical expression for the cross-correlation function of the process in the case of Gaussian pulses and thereafter calculate the time lag maximizing this cross-correlation function. The cross-correlation function is defined as
\begin{equation}
    R_{\Phi}(\xlag, \ylag, \tlag) = \ave{\Phi(x,y,t) \Phi(x+\xlag,y+\ylag,t + \tlag)}.
\end{equation}
Using \Eqref{eq.2d}, developing the product and regrouping the sum in terms with different indexes and the same index, we can write this as
\begin{align}
    &R_{\Phi}(\xlag, \ylag, \tlag) = \nonumber \\
    = &\ave{\sum_{k\neq j} \phi_k(x, y-\yk, t-\tk) \phi_j(x+\xlag, y-\yk+\ylag, t-\tk+\tlag)} + \nonumber \\
    &\ave{\sum_k \phi_k(x, y-\yk, t-\tk) \phi_k(x+\xlag, y-\yk+\ylag, t-\tk+\tlag)}
\end{align}
In the limit $T, L \rightarrow \infty$, the first term gives a contribution $\ave{\Phi}^2$ independent of both $\ylag$ and $\tlag$. For the second term, we can apply \Eqref{eq.avg.recepie} and change variables to the pulse coordinate system to obtain
\begin{align}
    &\ave{\sum_k \phi_k(x, y-\yk, t-\tk) \phi_k(x+\xlag, y-\yk+\ylag, t-\tk+\tlag)} = \nonumber \\
    &=\frac{\langle{a^2}\rangle\ellx\elly}{\rate v} \int_{-\infty}^{\infty} \rmd \tx \int_{-\infty}^{\infty} \rmd \ty  \varphi_\shortparallel (\tx) \varphi_\perp (\ty) \varphi_\shortparallel (\tx + \triangle_{\tx} ) \varphi_\perp (\ty + \triangle_{\ty}), 
\end{align}
where we adopt the shorthand notation $\tx = \tx(x, y-\yk, t-\tk)$ and $\ty = \ty(x, y-\yk, t-\tk)$ and we define
\begin{subequations}\label{eq.trit}
\begin{gather}
    \triangle_{\tx} = \frac{\xlag - v \tlag}{\ellx} \cos \tilt + \frac{\ylag - w \tlag}{\ellx} \sin \tilt, \label{eq.tritx.app}
    \\
    \triangle_{\ty} = -\frac{\xlag - v \tlag}{\elly} \sin \tilt + \frac{\ylag - w \tlag}{\elly} \cos \tilt. \label{eq.trity.app}
\end{gather} 
\end{subequations}
Now the integral can be written in terms of the pulse autocorrelation function given by \Eqref{eq.ps.acf} and the variance in \Eqref{eq.variance} to obtain
\begin{equation}\label{eq.ccf.general.app}
    R_{\Phi}(\xlag, \ylag, \tlag) = \ave{\Phi}^2 +\Phirms^2 \rho_\shortparallel(\triangle_{\tx})  \rho_\perp(\triangle_{\ty}).
\end{equation}
A further simplification can be made by considering the cross-correlation function of the normalized process $\wt{\Phi}$ instead, in which case it becomes
\begin{equation}\label{eq.ccf.general.norm.app}
    R_{\wt{\Phi}}(\xlag, \ylag, \tlag) = \frac{1}{I_{x, 2} I_{y, 2}} \rho_\shortparallel(\triangle_{\tx})  \rho_\perp(\triangle_{\ty}).
\end{equation}
Thus, in the case of Gaussian pulse functions given by \Eqref{eq.gauss}, the signal cross-correlation function becomes
\begin{equation}
    R_{\Phi}(\xlag, \ylag, \tlag) = \ave{\Phi}^2 + \Phirms^2\exp \left( -\frac{\triangle_{\tx}^2}{2}-\frac{\triangle_{\ty}^2}{2} \right),
\end{equation}
and the cross-correlation function of the normalized signal
\begin{equation}\label{eq.ccf.app}
    R_{\wt{\Phi}}(\xlag, \ylag, \tlag) = \exp \left( -\frac{\triangle_{\tx}^2}{2}-\frac{\triangle_{\ty}^2}{2} \right),
\end{equation}
where $\triangle_{\tx}$ and $\triangle_{\ty}$ are defined by \Eqsref{eq.tritx.app} and (\ref{eq.trity.app}). The time lag maximizing \Eqref{eq.ccf.app}, for an arbitrary tilt angle $\tilt$, is given by
\begin{align}\label{taumax.tilt}
&\tau_\text{max}(\xlag, \ylag)= \nonumber \\
&\frac{{(\xlag \elly^2 v + \ylag \ellx^2 w) \cos^2 \tilt - (\ellx^2 - \elly^2)(\ylag v + \xlag w) \cos \tilt \sin \tilt + (\xlag \ellx^2 v + \ylag \elly^2 w) \sin^2 \tilt}}{{(\elly^2 v^2 + \ellx^2 w^2) \cos^2 \tilt + \sin \tilt(-2(\ellx^2 - \elly^2)vw \cos \tilt + (\ellx^2 v^2 + \elly^2 w^2) \sin \tilt)}}.
\end{align}
It can also be written in terms of $\beta=\arctan w/v$ and $u = \sqrt{v^2+w^2}$ as
\begin{align}\label{taumax.tilt.angles}
&\tau_\text{max}(\xlag, \ylag)= \nonumber \\
&\frac{\xlag(\elly^2-\ellx^2)\cos(2\tilt-\vdir)+\xlag(\ellx^2+\elly^2)\cos\vdir + \ylag(\elly^2-\ellx^2)\sin(2\tilt-\vdir)+\ylag(\ellx^2+\elly^2)\sin\vdir}{u(\ellx^2 + \elly^2+(\elly^2-\ellx^2)\cos[2(\tilt-\vdir)])}.
\end{align}
Equations~\eqref{taumax.tilt} and (\ref{taumax.tilt.angles}) give a general expression for the time maximizing the cross-correlation function resulting from a superposition of elongated pulses with arbitrary velocities. These expressions simplify in the case that $\ellx=\elly$, as then the system becomes independent of the tilt angle $\tilt$. More importantly, in the case that the tilt angle $\tilt$ aligns with the pulse propagation $\tilt=\vdir$, $\tau_\text{max}(\xlag, \ylag)$ becomes
\begin{equation}
    \tau_\text{max}(\xlag, \ylag) = \frac{1}{u}\left( \xlag\cos\vdir + \ylag\sin\vdir \right) = \frac{\xlag v + \ylag w}{v^2 + w^2} .
\end{equation}
This result applies even if $\ellx\neq\elly$, showing that the time lag maximizing the cross-correlation function becomes independent of the sizes $\ellx$ and $\elly$ when $\tilt=\vdir$. This is the result used in \Secref{sec.estimation.aligned}.

\section{Alternative pulse functions}\label{app.ps}

\begin{figure}[tb]
\centering
\includegraphics[width=\columnwidth]{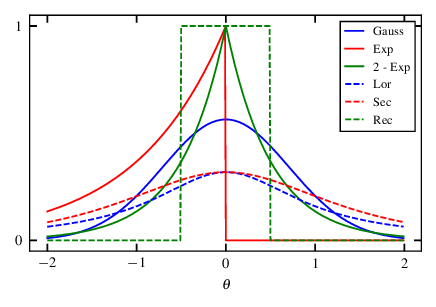}
\caption{Commonly used pulse functions defined in Table~\ref{table.ps}.}
\label{fig.pulse_functions}
\end{figure}

The velocity estimation method studied in this contribution has been derived analytically for Gaussian pulse functions. There is no theoretical reason nor experimental observation that justifies the assumption of Gaussian pulses other than mathematical simplicity. Therefore, it is interesting to consider the robustness of the method in the case of other pulse functions. A list of some commonly used analytical pulse functions, together with their autocorrelation functions, is given in Table~\ref{table.ps} and shown in \Figref{fig.pulse_functions}. These are Gaussian, one-sided exponential, two-sided exponential, Lorentzian, hyperbolic secant and rectangle pulses. For simplicity, we will assume that the parallel and perpendicular pulse functions are the same.

\begin{figure*}[tb]
\centering
\includegraphics[width=\textwidth]{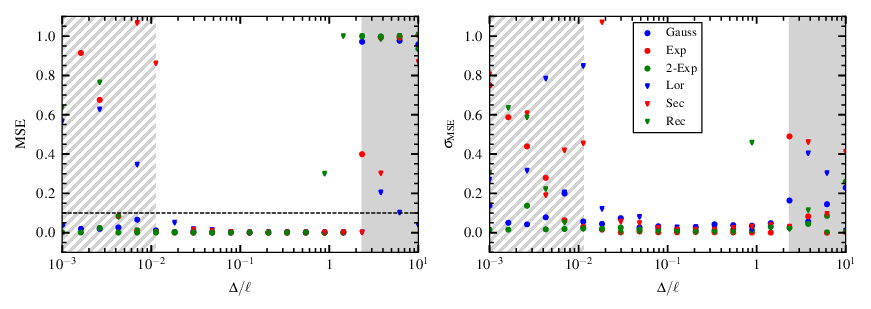}
\caption{Left: Mean square error of the three-point velocity estimation for realizations of the process with different pulse functions. Each plot symbol represents the average mean-square error of 10 independent realizations. The gray-shaded region on the right represents the parameter space where the cross-correlation maxima fall under $0.25$. The hatched region on the left represents the parameter space where the maximum estimated time delay is lower than the sampling time. Right: Mean square error standard deviation over the $10$ simulations.}
\label{fig.pulse_functions_res}
\end{figure*}

We performed a series of simulations for each pulse function with changing spatial resolution. The velocities are estimated using the method derived for Gaussian pulse functions. The results from the three-point velocity estimations are presented in \Figref{fig.pulse_functions_res}. The gray-shaded region on the right represents the parameter space where the cross-correlation maxima fall under $0.25$. This cross-correlation decays differently for different pulse shapes, the gray-shaded region is computed from the average cross-correlation. The rectangle and exponential pulse function leads to signals where the cross-correlation decays fastest with increasing $\Delta/\ell$, followed by two-sided exponential (2-Exp), Gaussian, Lorentzian and secant. In general, no large deviations in the estimation errors are observed with respect to the Gaussian case. The largest deviation is found in the case of rectangular pulse functions, likely due to their compact support.

\begin{center}
\begin{table}[tb]
\begin{tabular}{ |c |c| c |}
\hline
 Pulse function & $\varphi(\theta)$ & $\rho_\varphi(\Delta \theta)$ \\
\hline
 Gaussian & $\frac{1}{\sqrt{\pi}} \exp (-\theta^2) $ & $\frac{1}{\sqrt{2\pi}} \exp \left( -\frac{ \Delta \theta^2 }{2}\right)$ \\ 
 Exp & $\exp (\theta) \Theta(-\theta)$ & $\frac{\exp(-\abs{\Delta \theta})}{2}$ \\  
 2-Exp & $\exp \left( -2|\theta| \right)$ & $\exp \left( -2|\theta|\right)$    \\
 Lorentz & $\frac{1}{\pi\left( 1 + \theta^2\right)}$ & $\frac{2}{\pi \left( 4 + \Delta \theta^2\right)}$ \\
 Sech & $\frac{2}{\sqrt{\pi} \left(\exp (\theta) + \exp(-\theta)\right)}$ & $\frac{4 \Delta \theta}{\pi^2 \left( \exp(\Delta \theta) - \exp(-\Delta \theta)\right)}$    \\
 Rec & $\Theta \left( 1/2 - |\theta| \right)$ & $(1-|\theta|) \Theta \left( 1 - |\theta| \right)$\\
\hline
\end{tabular}
\caption{Commonly used pulse functions together with their autocorrelation function.}
\label{table.ps}
\end{table}
\end{center}

\section{Barberpole effect on the two-point method}\label{app.2pbp}

Here we study some properties of the two-point velocity estimation method. This technique estimates the velocity components from the time delays in the following way:
\begin{subequations}\label{eq.2tde}
\begin{align}
    \wh{v}_2 & = \frac{\dx}{\tau_x},
    \\
    \wh{w}_2 & = \frac{\dy}{\tau_y},
\end{align}
\end{subequations}
where $\dx$, $\dy$, $\tau_x$ and $\tau_y$ are defined in \Secref{sec.estimation}. In general, this method will give wrong estimates for an arbitrary velocity direction. In the case that pulses are tilted in the direction of propagation, $\tilt=\vdir$, it can be shown that the two-point method estimates deviate by a factor $\wh{v}_2/v=1+w^2/v^2$ and $\wh{w}_2/w=1+v^2/w^2$ \cite{losada_three-point_2024}. Only in the case that the velocity is aligned with one of the separations from which $\tau_x$ and $\tau_y$ are estimated, will this method provide a good estimate for that velocity component.

\begin{figure*}[tb]
\centering
\includegraphics[width=\textwidth]{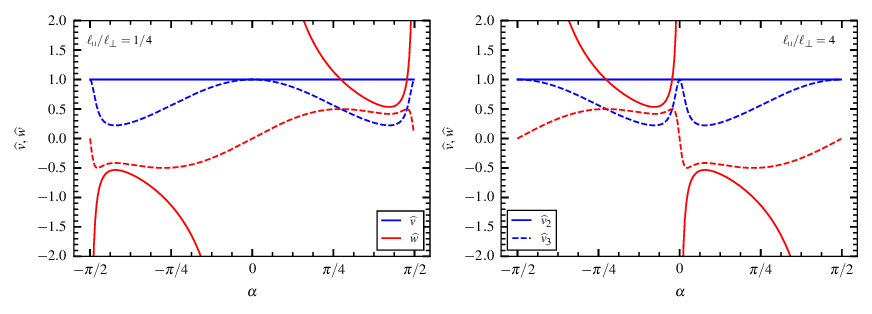}
\caption{Analytical functions of the two-point (solid lines) and three-point (dashed lines) methods for the estimates of the horizontal (blue) and vertical (red) velocity components. The true velocity components are $v=1$ and $w=0$. The results are shown for elongations $\ellx/\elly=1/4$ (left) and $\ellx/\elly=4$ (right).}
\label{fig.analytical_barberpole}
\end{figure*}

Here we are interested in quantifying the deviation of these estimates from the true velocities in the case of elongated and tilted pulses, that is, when the barberpole effect is present. For a given ratio $\ellx/\elly$ and assuming $\dx=\dy=\Delta$ as before, this deviation can be computed analytically with \Eqref{taumax.tilt}. The results are shown in \Figref{fig.analytical_barberpole} for the two-point method (solid lines) and the three-point method (dashed lines) in the case that the true velocities are $v=1$ and $w=0$ for $\ellx/\elly=1/4$ and $4$. The curves for the three-point method are equivalent to those presented in \Figref{fig.tilt}. Note that the two-point method always gives the correct estimate for the horizontal velocity components $\wh{v}$. This is expected as the true velocity is indeed only horizontal. The barberpole effect does not affect this estimate. On the other hand, the vertical component, which would be estimated to be infinity $\wh{w}_2=\infty$ in the absence of a barberpole effect, is now finite for $\alpha\neq 0$. Both for the two- and three-point methods, the curves for the case $\ellx/\elly=1/4$ are displaced at an angle $\pm\pi/2$ with respect to the case $\ellx/\elly=4$. Indeed, a pulse with $\ellx/\elly=1/4$ and $\alpha=0$ is equivalent to a pulse with $\ellx/\elly=4$ and $\alpha=\pm\pi/2$.
These results showcase an interesting property of the two-point method: if prior knowledge of the system can guarantee that the pulses move parallel to the direction separating two measurement points, the two-point method will estimate the correct velocity even in the presence of the barberpole effect.

\bibliographystyle{apsrev4-1}
\bibliography{SOL}

\end{document}